# Long period slow MHD waves in the solar wind source region

B.N. Dwivedi [*], A.K. Srivastava[**]

*Department of Applied Physics, Institute of Technology, Banaras Hindu University, Varanasi-221 005, India*

______________________________________________________________________

**Abstract**

We consider compressive viscosity and thermal conductivity to study the propagation and dissipation of long period slow longitudinal MHD waves in polar coronal holes. We discuss their likely role in the line profile narrowing, and in the energy budget for coronal holes and the solar wind. We compare the contribution of longitudinal MHD waves with high frequency Alfvén waves.

*PACS:* 96.50.Tf; 96.60.pc; 96.60.Vg

*Keywords*: MHD waves; solar wind; coronal holes.

*Corresponding author

*E-mail address*: bholadwivedi@gmail.com (B.N. Dwivedi)

** Present Address: *Armagh Observatory, College Hill, Northern Ireland BT61 9DG.*

______________________________________________________________________

## 1. Introduction

The broadening of line profile with radial height is the likely signature of the undamped Alfvén wave propagation (Hassler et al., 1990; Wilhelm et al., 2004, 2005; and references cited therein). The narrowing of line profile is, however, attributed to the propagating damped Alfvén waves (Harrison et al., 2002; O'Shea et al., 2005). Spectroscopic studies, both on the disk and off-limb, also show the presence of compressional waves in plumes and inter-plume regions of polar coronal holes and their possible role in the solar wind acceleration (Gurman and DeForest, 1998; Ofman et al.,1997, 1998; Banerjee et al., 2001). Recently, O'Shea et al. (2006, 2007) have also inferred the presence of fast and slow magnetoacoustic waves in the polar off-limb region. Linear MHD model of Alfvén wave dissipation in the equatorial corona has been reported by Harrison et al. (2002). The same explanation is given by Pekünlü et al. (2002)

in the northern polar coronal holes, invoking linear MHD model correlating with the observed line profile narrowing in these regions. Dwivedi and Srivastava (2006) have also studied the outwardly propagating Alfvén waves and their dissipation in coronal holes, supporting the solar wind outflow starting between 5 Mm and 20 Mm as reported by Tu et al. (2005 a, b). The quasi-periodic disturbances in the off-limb above the polar coronal holes have been interpreted as the slow magnetosonic waves by Ofman et al. (2000b). Using Extreme-Ultraviolet Imaging Telescope (EIT) on board the SOHO spacecraft, Ofman et al. (1999) have also found the signatures of quasi-periodic compressional waves with periods of 10–15 minutes in southern polar plumes, and interpreted them as slow magnetosonic waves. These waves propagate with the subsonic speed of 75–150 km s$^{-1}$ and steepen in the corona nonlinearly. The steepening may lead to the formation of shocks higher in the corona, and may contribute significantly to the solar wind acceleration by transferring momentum to the flow. Ofman et al. (2000 a) have found that nonlinear steepening of the slow magnetosonic waves leads to enhanced dissipation owing to compressive viscosity at the wave fronts. The efficient dissipation of the slow wave by compressive viscosity leads to damping of the waves within the first solar radii above the limb. The spherically symmetric Alfvén waves are also subjected to the nonlinear steepening as they move up in the higher corona (Nakariakov et al., 2000). Suzuki and Inutsuka (2005) have also found that after the attenuation of the 85% of the initial energy flux in the chromosphere, the outgoing Alfvén waves enter the corona and contribute to the heating and acceleration of the plasma mainly by the nonlinear generation of compressive waves and shocks. Hence, the study of both linear and nonlinear MHD waves are important to describe the heating of coronal holes and acceleration of the fast solar wind. The signature of slow wave oscillations and its dissipation has also been studied in hot coronal loops by Ofman and Wang (2002). Mendoza-Briceno et al. (2002, 2005) have also studied numerically the response of the coronal plasma to microscale heating pulses in loops.

In this paper, we consider thermal conductivity and compressive viscosity as dissipating agents to investigate slow longitudinal MHD waves with periods 60 s, 180 s, and 300 s respectively. We consider the solar wind source region and its upper part inside polar coronal holes for the propagation of slow longitudinal MHD waves along the background magnetic field direction. Our first aim is to explore whether or not the dissipation of MHD waves causes observed line profile narrowing (O'Shea et al., 2005). In the light of spectroscopic signature of line-width variation in polar coronal holes, our second aim is to compare the role of high-frequency Alfvén waves (Srivastava and Dwivedi, 2007; Dwivedi and Srivastava, 2006) with the slow longitudinal MHD waves in the inner corona. The nonlinearity and background plasma flow can affect the properties of these MHD waves, e.g., cut-off frequencies, dissipation rates, and phase speeds etc (e.g., Terra-Homem et al., 2003). However, we consider the linear approximation in order to explore our primary goal of explaining the observation. We estimate energy flux density of MHD waves in the linear approximation to calculate theoretical line width. Since observed spectral line width variation is along outward radial direction only, we consider only the radial outward propagation of MHD waves, neglecting resonant absorption and phase mixing. In Section 2, we describe MHD equations and theoretical considerations. In Section 3, we describe theoretical line widths, estimated as a result of Alfvén waves, and slow longitudinal MHD waves in polar coronal holes. Results and discussion are presented in the last Section.

## 2. MHD equations and theoretical considerations

We consider propagation of slow longitudinal MHD waves along the background magnetic field in polar coronal holes. The background magnetic field is considered along z-axis, which is the normal outward direction of the Sun's surface. We study the combined effect of compressive viscosity and thermal conductivity on the propagation and dissipation of these waves in polar coronal holes. One-dimensional MHD equations for viscous and thermal conductive plasma are taken from De Moortel and Hood (2003) (cf., Eqs. 1-4).

Compressive viscosity and thermal conductivity are: $h_o = 10^{-16} T_o^{2.5}$ (g cm$^{-1}$ s$^{-1}$) and $k_\parallel = 10^{-6} T_o^{2.5}$ (ergs cm$^{-1}$ s$^{-1}$ K$^{-1}$) (Braginskii 1965). In equilibrium, plasma pressure, temperature and density respectively are $p_o$, $T_o$ and $r_o$. The other symbols have their same meanings as given in De Moortel and Hood (2003). Using equilibrium values of pressure, temperature, and density, they made their Eqs. (1) - (4) dimensionless. The velocity parameter is made dimensionless by dividing it with adiabatic sound speed, $c^2{}_s = \gamma p_0 / \rho_0$. The length and time are made dimensionless by particular plasma column length L, and observed time period $t$, which are related as $L = c_s \tau$. The linearized equations with dimensionless parameters are also taken from De Moortel and Hood (2003) (cf., Eqs. 9-12). The two dimensionless parameters $e = \dfrac{h_o t}{r_o L^2} = \dfrac{h_o}{g t p_o}$, and $d = \dfrac{(g-1) k_\parallel T_o r_o}{g^2 p_o^2 t}$ are also similar to their Eqs. 7 and 8 respectively. The details of parameters $e$ and $d$ are given in De Moortel and Hood (2003).

Assuming all the variables having the phasor factor $\exp(i k z - i w t)$ and solving linearized equations simultaneously, we get the following dispersion relation:

$$A k^6 + B k^4 + C k^2 + D = 0 \qquad (1)$$

where

$$A = \frac{4}{3} i g^2 e d + \frac{16}{9} g^3 e^2 w d ,$$

$$B = g^2 w d + \frac{4}{3} g^2 w e - \frac{8}{3} i g^3 w^2 e d - \frac{16}{9} i g^2 w^2 e^2$$

$$C = -g^3 w^3 d - \frac{8}{3} g^2 w^3 e - i w^2 g^2 ,$$

and

$$D = i g^2 w^4 .$$

We take the density profile in a polar coronal hole as a function of radial height from the empirical relation of Doyle et al. (1999) :

$$N_e(R) = \frac{1 \times 10^8}{R^8} + \frac{2.5 \times 10^3}{R^4} + \frac{2.9 \times 10^5}{R^2} \quad (\text{cm}^{-3}) \qquad (2)$$

The mass density is $r_o(R) = 0.6 m_p N_e(R)$, where $m_p$ is the proton mass. The empirical relation of temperature profile is obtained using Habbal et al. (1993) temperature measurements :

$$T_R = (-10.51235 + 23.5623 R - 15.77 R^2 + 3.53147 R^3) \times 10^6 \text{ (K)}. \qquad (3)$$

The energy flux density of MHD wave is given by

$$W = r v_{NT}^2(R) V , \qquad (\text{ergs cm}^{-2} \text{ s}^{-1}) \qquad (4),$$

where $V$ is wave velocity and $v_{NT}(R)$ is velocity equivalent to the non-thermal component of relevant spectral line at FWHM. We use the empirical relation for $v_{NT}(R)$ as a function of radial height, given by Pekünlü et al. (2002) using Banerjee et al.(1998) measurements for Si VIII ion :

$$v_{NT}(R) = -1522.3R^4 + 8638.2R^3 - 18191R^2 + 16882R - 5786.5 \ (\text{km s}^{-1}). \qquad (5)$$

The data below 1.02 $R_\odot$ and above 1.26 $R_\odot$ have been obtained by extrapolation. . We get 22.24 km/s at 1.0058 $R_\odot$ , 28.96 km/s at 1.03 $R_\odot$, 47.6 km/s at 1.25 $R_\odot$, and 47.8 km/s at 1.35 $R_\odot$. Hence, we can see that these values of nonthermal velocity are reasonable in the region of interest (1.0058 $R_\odot$- 1.35 $R_\odot$).

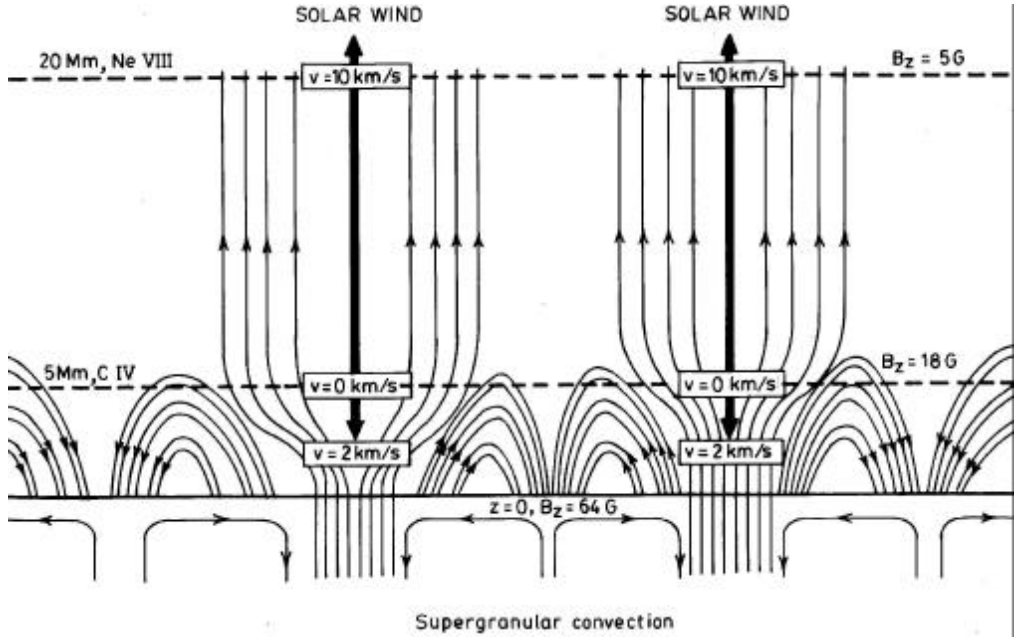

Fig.1. Schematic model of the solar wind source region (magnetic funnels), and its energy and mass supply from side loops, adapted from Tu et al. (2005 b).

Tu et al. (2005 a, b) have suggested a new model, which illustrates the scenario of the solar wind origin in coronal funnels, and its energy and mass supply from side loops. We have adapted their idea to make a schematic representation of side loops and funnels which is shown in Fig. 1. Previously, the solar wind was believed to originate from the ionized hydrogen layer slightly above the photosphere. However, Tu et al. (2005 a) have shown from the Doppler shift of C IV emission line that the bulk flow of the solar wind did not occur at a height of 5 Mm. They determined the height of the solar wind origin in the magnetic funnels by a comparative and correlative study of coronal magnetic field obtained from the force-free extrapolation of the photospheric field using the MDI data, and the radiance and Doppler maps of EUV spectral lines (e.g., Si $^{II}$, C $^{IV}$, and Ne $^{VIII}$) as observed by the SUMER/SOHO spectrograph. They found that C $^{IV}$ intensity has a maximum correlation coefficient with the extrapolated magnetic field and does not show a significant Doppler shift at a height of about 5 Mm. But C $^{IV}$ lines may have a blue-shift near the centre of the supergranular cells and is partially red-shifted near the cell boundaries. Tu et al. (2005 a, b) have also found the blue Doppler shift for the Ne $^{VIII}$ ion at a height of 20.6 Mm in the funnels. The transition region of coronal holes is full of bipolar loops up to a maximum height of 7 Mm. The neighbouring loops keep the funnels' cross-section constricted and prevent a rapid horizontal expansion through magnetic tension to the funnels with height. The adjacent magnetic loops are the likely carriers of energy to the solar wind plsama. We consider the solar wind source region (i.e., 1.0058 $R_\odot$ - 1.03 $R_\odot$) and its upper part (i.e., 1.03 $R_\odot$ - 1.35 $R_\odot$) of polar coronal holes, for studying the propagation and dissipation of long period slow longitudinal MHD waves. Popescu et al. (2004) also inferred the observed blue-shift in coronal lines (e.g., Mg IX 706 Å line), and an average red-shift in lower transition region lines (e.g., O III 703 Å line). However, they found that the red-shifted appearance of this line is sprinkled with blue-shifts forming a small-scale network pattern with an average outflow speed of 3.5 km s$^{-1}$ in coronal hole. However, this analysis does not provide the height information of the solar wind origin. If we consider that O III emits near 3 Mm height in solar atmosphere (Doschek et al., 1976), we may consider the sprinkles of the outflow of plasma around this height. However, the more sophisticated three-dimensional picture of the solar wind source region and coronal magnetic field is firstly given by Tu et al. (2005 a,b).

We solve the dispersion relation (Eq. 1) which provides six roots to find out damping length scale, energy flux density, and wavelength etc. Only two complex roots, and parameters derived therefrom are important. They represent the slow longitudinal MHD waves in the solar atmosphere. Hence, we present the results which are derived from these two roots. We consider slow longitudinal MHD waves of periods 60 s, 180 s and 300 s in our calculation.

## 3. Line widths: Alfvén waves and slow longitudinal MHD waves

Alfvén waves are the transversal and non-compressional waves which do not modulate the plasma density, in contrast to slow mode acoustic waves and fast mode compressional MHD waves which do modulate the plasma density. The fast mode MHD waves modulate the plasma density to a lesser extent in comparison to slow mode acoustic waves. Hence, the study of periodicity in the EUV light curve is the useful way to detect the compressional modes from the observations. However, both the slow acoustic waves and Alfvén waves perturb the plasma velocity which causes positive and negative Doppler shift that can be detected as a line width broadening or line width variation. Hence, Alfvén waves may be a primary candidate for the line width broadening or narrowing and can be detected by studying the line width (e.g., Banerjee et al., 1998, Doyle et al., 1998, Moran, 2003). However, the slow waves which are clearly present in the inner part of polar coronal holes, may also contribute to the variation of line width. The Doppler line widths as a result of radially outward propagating Alfvén waves, and slow longitudinal MHD waves are given as follows (e.g., Erdélyi et al., 1998) :

$$\Delta l_{D\,Alf} = \frac{1}{c}\left(\frac{2kT}{M_i} + \frac{4\,p^{0.5}\,f_{Alf}}{r^{0.5}\,B}\right)^{0.5} l \qquad (6),$$

and

$$\Delta l_{D\,Lon} = \frac{1}{c}\left(\frac{2kT}{M_i} + \frac{2\,f_{Lon}}{N_e\,(6m_p kT)^{0.5}}\right)^{0.5} l \quad. \qquad (7)$$

We take $c = 3 \times 10^{10}$ cm s$^{-1}$, $k = 1.38 \times 10^{-16}$ ergs K$^{-1}$, $M_i = 4.008 \times 10^{-24}$ g for Mg $^X$ ion, $l = 609.78$ Å, proton mass $m_p = 1.67 \times 10^{-24}$ g. The magnetic field is taken from the empirical relation of Dwivedi and Srivastava (2006),

$$B = 95.20344 - 129.39079\,R + 45.49857\,R^2, \text{ for } 1.05\,R_\odot < R < 1.35\,R_\odot. \qquad (8)$$

$f_{Alf}$ is the energy flux density for the Alfvén waves (Dwivedi and Srivastava, 2006), and $f_{Lon}$ is the energy flux density of slow longitudinal MHD waves, which are calculated in this paper.

## 4. Results and Discussion

Fig. 2 shows the dependence of thermal conductivity and compressive viscosity on the spatial variation of wavelength of slow longitudinal MHD waves with time periods 60 s, 180 s, and 300 s respectively. For a particular wave period, wavelength of the two slow modes exhibits increasing pattern with height above the limb. Our dispersion relation is valid for the description of the waves whose wavelengths are shorter than the density and magnetic field stratification height. The density scale height in the inner corona approximately ranges between 50 and 60 Mm. Our theoretically estimated wavelength lies in the range between 5 Mm and 40 Mm. Hence, the wavelength is shorter than the density scale height in the lower corona, which validates the use of our dispersion relation for such waves. Assuming Alfvén speed $1 \times 10^3$ km s$^{-1}$ and sound speed $0.2 \times 10^3$ km s$^{-1}$ in the inner corona, our dispersion relation is valid for the slow longitudinal MHD waves with maximum cut-off period of 300 s. The wavelengths for different wave periods are slightly different with each other at any particular height. For example, the difference between the wavelengths of 300 s and 60 s waves is ~ 18 km at 1.35 $R_\odot$. The table for wavelengths is attached in the Appendix for the sake of clarity. The Y-scale in Fig. 2 being of the order of ~ $10^4$ km, we are not able to resolve small differences in the wavelengths at different periods. However, Table I appended to this paper clearly shows the dependence of wavelength on wave-period, thereby being physically valid.

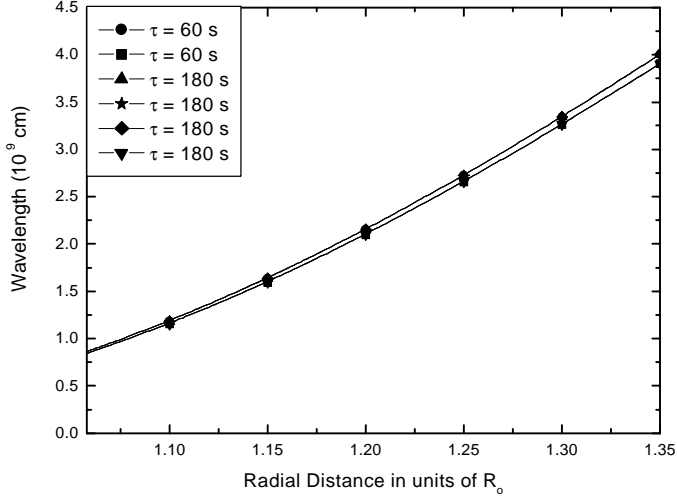

Fig.2. Spatial variation of wavelength of slow longitudinal MHD waves with the periods 60s, 180s, and 300s.

Fig. 3 shows the dependence of thermal conductivity and compressive viscosity on the spatial variation of damping length scale of slow longitudinal MHD waves for time periods 60 s, 180 s, and 300 s respectively. For a particular wave period, damping length scale of two slow modes exhibits increasing pattern with height above the limb. Hence, we see that damping through compressive viscosity and thermal conductivity is rather weak as we move upward. The wavelength and the nature of its variation with radial height being approximately comparable to the damping length scale, validate the linear MHD model in our study. The damping length scales for different wave periods are slightly different with each other at any particular height. For example, the difference between the damping length scales of 300 s and 60 s waves is ~ 3 km at 1.35 $R_\odot$. The Y-scale in the Fig. 3 is of the order of ~ $10^3$ km, hence, we are not able to resolve the small differences in the damping length scale at different periods. However, the appended Table II clearly shows the dependence of damping length scale on wave-period, therefore, physically valid.

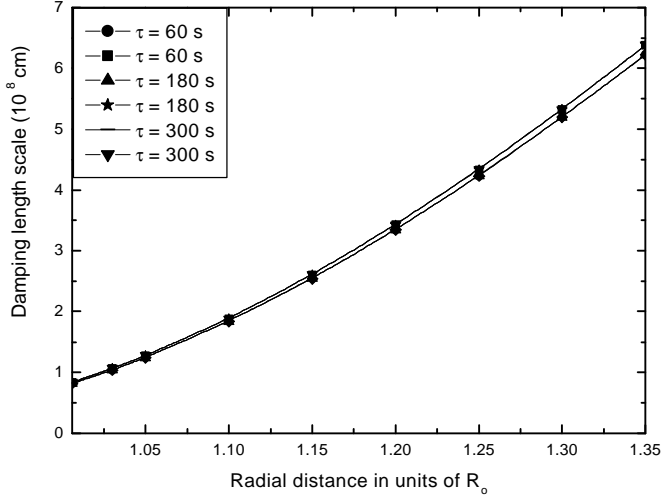

Fig. 3. Spatial variation of damping length scale of slow longitudinal MHD waves with the periods 60s, 180s, and 300s.

Fig. 4 shows the combined effect of thermal conductivity and compressive viscosity on the spatial variation of the energy flux density of slow longitudinal MHD waves for time periods 60 s, 180 s, and 300 s respectively. For a particular time period, the energy flux density of the two slow modes first increases up to a height of ~ 1.21 $R_\odot$ and then decreases. The energy flux density is not necessarily to be conserved in the dissipative medium (Zaqarashvili, 2007). The conservation of energy flux density generally holds when there is neither dissipation effects e.g., viscosity, thermal conductivity, resistivity, ion-neutral collisions etc., nor 3-D non-linear effects (Nakariakov, 2007). The other aspect is the uncertainty in the observational parameters, such as density, temperature, nonthermal velocity, used in the theoretical calculations. The most crucial factor is the nonthermal velocity derived from observed line width, which may contain the contribution of different nonthermal energy sources other than slow waves. While the theoretically calculated wave velocity is the velocity of slow waves, the energy flux density may not necessarily be conserved with height (Ofman, 2007). The uncertainty in the observed parameters introduces some degree of uncertainty in the profile of energy flux density, which may result in the growth of energy flux density initially. However, we note the importance of the location (~ 1.21 $R_\odot$) where the dissipation starts, and from where and beyond slow waves may also contribute to the observed line width reduction. Every simple or complex model aims at exploring new physics under its physically viable assumptions and constraints. This may result in uncertainties in the computation of energy flux density and in other parameters in our model, while exploring some important aspects of coronal physics. O'Shea et al. (2005) measurements of Mg $^X$ 609.78 Å and 624.78 Å lines from Coronal Diagnostic Spectrometer (CDS), provide the variation of line widths and line ratio in the regions far off-limb in northern polar coronal holes. They have found a decrement in the line widths above ~ 1.21 $R_\odot$ and have attributed it to a change of collisionally-excited plasma into the radiatively-excited one. Using the line width measurement, Dwivedi and Srivastava (2006) have calculated the non-thermal velocity variation in the inner part of polar coronal holes. They have calculated Alfvén wave energy flux density that shows a decrement beyond ~1.21 $R_\odot$ where non-thermal velocity, deduced from observations, also starts reducing. These results indicate the dissipation of outwardly propagating Alfvén waves, which cause the reduction in non-thermal component of the observed line widths. These results also show that Alfvén waves of period 0.001 s fit the observational data best. In the case of slow longitudinal MHD wave, the energy flux density also shows a decrement at the same location. We interpret this result as a dissipation of these waves beyond ~ 1.21 $R_\odot$. This result is also supported by the reduction of the energy flux density of slow longitudinal MHD waves for all time periods and their peaks also match with the peaks in the radial profiles of observed line widths and non-thermal velocity. It seems that not only Alfvén waves, but also dissipation of slow longitudinal MHD waves are responsible for the decrement in the line width and nonthermal velocity at the height ~ 1.21 $R_\odot$. The energy flux density of 60 s wave is of the order of ~ $10^4$ ergs cm$^{-2}$ s$^{-1}$ in coronal hole and solar wind, however, it is of the order of ~ $10^3$ ergs cm$^{-2}$ s$^{-1}$ for the slow longitudinal MHD waves with periods 180 s and 300 s. Doyle et al. (1998) have reported that purely acoustic waves are not too important for coronal heating (Hollweg, 1990), and the slow and fast waves are not able to supply the coronal energy requirements. There may be some uncertainty in the measurements of temperature, density, magnetic field, and hence in the estimation of wave energy flux density also. However, we can hardly ignore the partial role of these waves for coronal hole heating and the solar wind acceleration. In any case, our study shows the importance of location near ~ 1.21 $R_\odot$, where both Alfvén as well as longitudinal waves exchange energy with the coronal holes and solar wind via collisional dissipation, and cause the observed line profile narrowing.

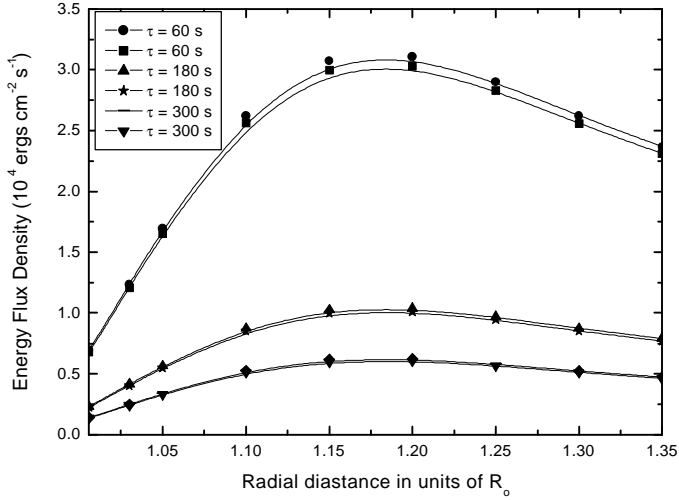

Fig. 4. Spatial variation of energy flux density of slow longitudinal MHD waves with the periods 60s, 180s, and 300s. We have also examined the role of dissipating agents individually, noting that the effect of thermal conductivity is larger in comparison to viscosity.

Using Eqs. (6) and (8) respectively, we estimated Mg $^X$ 609.78 Å line width for Alfvén and slow longitudinal MHD waves in polar coronal holes, and compared them with observation reported by O'Shea et al. (2005). Fig. 5 shows that Mg $^X$ 609.78 Å line width shows a decrement at ~1.21 $R_\odot$. Our theoretically estimated line widths as a result of Alfvén wave and long period slow MHD waves of periods 60 s, 180 s, and 300 s are respectively shown in Fig. 5. Although theoretical curves in Fig. 5 deviate, yet they show approximately the same nature as observed. These theoretical curves also show decreasing or flat arms beyond ~ 1.21 $R_\odot$. The theoretically calculated line width curves become flat in case of both slow longitudinal MHD and Alfvén wave beyond ~ 1.21 $R_\odot$ and finally show the decrement in the case of Alfvén wave. The increment rate of these curves decreases after 1.20 $R_\odot$. It is obvious that theoretically calculated values cannot trace the observations exactly, due to the uncertainty of the physical parameters. This behaviour of the theoretical curves can be inferred as the reduction in the non-thermal component of the line width due to the MHD wave dissipation. It is also interesting to note that the curve for 60 s slow longitudinal MHD wave is more close and best fit with observations at 1.20 $R_\odot$ and beyond this height. This fits the observation more closely than Alfvén wave with period 0.001 s. This favours the importance of long period slow MHD waves with period 60 s at the location 1.20 $R_\odot$ and beyond this height. We tend to suggest that 60 s slow MHD wave contributes more to the line profile narrowing through dissipation. The deviation in the line width values as estimated by theory and observation, are due to large uncertainties in energy flux density of MHD waves, density, temperature, and magnetic field topology of the solar corona. Hence, we cannot use the theoretically calculated values to exactly fit the observational data. This clearly shows the correlation between the observed line width and the dissipation of both types of waves. Erdélyi et al. (1998) have studied the center to limb line width variation of the upper chromospheric and transition region lines. Their numerical studies favour the Alfvén wave heating over the magnetoacoustic heating. Their calculated theoretical line width variations due to Alfvén wave fit the observational data best, however, this is not the case for the magnetoacoustic waves. When we estimated theoretical line widths as a result of both types of waves in polar coronal holes, we find that both slow longitudinal MHD waves and Alfvén waves are important in this region and favour the observations qualitatively.

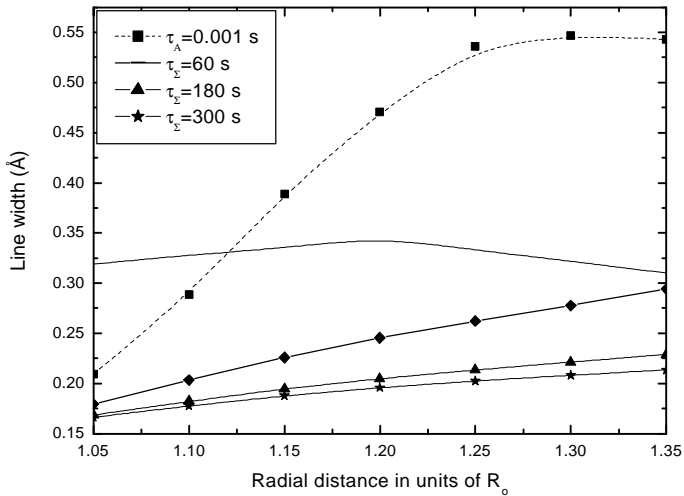

Fig. 5. Spatial variation of theoretical and observational Mg X 609 Å line widths. The solid curve shows the observed line width, while the lines with symbols show theoretical line widths.

The present study shows that energy flux densities of long period slow MHD waves in the solar wind source region are too low. Hence, we can only account the major role of high frequency Alfvén waves in the solar wind source region (1.0058 $R_\odot$ - 1.03 $R_\odot$). Alfvén waves are likely candidates in the solar wind source region. The energy flux of the long period slow longitudinal MHD waves is of the order of ~ $10^4$ ergs cm$^{-2}$ s$^{-1}$ in upper part of this region (1.03 $R_\odot$ - 1.35 $R_\odot$). Hence, they may be one of the possible energy sources in this part, and may contribute a fraction of energy required for coronal hole and solar wind.

The nonlinear effects and background plasma flows are also important in the study of propagation properties and resultant coronal heating. This nonlinearity and background plasma flow may affect the wave properties and also the observed line width, if it is thought that non-thermal energy sources contribute in the formation of the width of spectral lines. The effect of background plasma flow and flow modified resonant absorption may greatly affect the MHD wave (e.g., Alfvén waves) dissipation in coronal holes. The mathematical formulation, physical significance, and the effects of such processes are discussed in Erdélyi et al. (1995, 1996) and Doyle et al. (1997). However, the limitations of our 1-D MHD model (choice of the equations) constrain us to directly take account of these significant alternative dissipation mechanisms in our present study. We have aimed at explaining the relative contribution of Alfvén waves and slow waves, which dissipate through primary collisional dissipating processes (viscosity and thermal conductivity), in observed line profile narrowing in coronal holes. Moreover, we compare the role of these waves in the solar wind source region. We have chosen the initial equations in our study keeping in mind the above mentioned two very significant objectives. In the first instance, we are primarily successful in the completion of our objectives. However, this will be very significant and interesting subject of future work, if one can study the effect of alternative heating mechanisms and processes on the observed line profile narrowing by using the models as discussed in above mentioned references. This will also be interesting to study the difference between classical dissipating agents (viscosity, thermal conductivity, resistivity etc., as we did in the present work) and alternative but significant dissipative processes (mass flow, resonant absorption etc., in explaining the observations).

In conclusion, our study underlines the importance of outwardly propagating slow longitudinal MHD waves and their dissipation through viscosity and thermal conductivity in the inner polar coronal holes, supporting their role in the observed line width narrowing.

**Acknowledgements**

This work was supported by the Indian Space Research Organization under its RESPOND programme. We wish to express our gratitude to all the three anonymous referees for their valuable comments which improved the manuscript considerably. AKS gratefully thank V.M. Nakariakov, L. Ofman and T. Zaqarashvili, for their valuable suggestions during the revision of this paper. We also wish to thank the Editor, Dr. W. Soon for encouragements.

# *Appendix*

Table I: Wavelengths

| Radial Distance in units of $R_\odot$ | Wavelength for period 60 s (in $10^9$ cm) | Wavelength for period 60 s (in $10^9$ cm) | Wavelength for period 180 s (in $10^9$ cm) | Wavelength for period 180 s (in $10^9$ cm) | Wavelength for period 300 s (in $10^9$ cm) | Wavelength for period 300 s (in $10^9$ cm) |
|---|---|---|---|---|---|---|
| 1.0058 | 0.52612 | 0.51321 | 0.52595 | 0.51337 | 0.52618 | 0.51315 |
| 1.03 | 0.66822 | 0.65238 | 0.66861 | 0.65204 | 0.66863 | 0.65201 |
| 1.05 | 0.80004 | 0.78023 | 0.79999 | 0.78026 | 0.79982 | 0.78041 |
| 1.10 | 1.1806 | 1.1533 | 1.181 | 1.153 | 1.1817 | 1.1523 |
| 1.15 | 1.6342 | 1.5942 | 1.6338 | 1.5946 | 1.6349 | 1.5936 |
| 1.20 | 2.1514 | 2.0977 | 2.1516 | 2.0975 | 2.1508 | 2.0982 |
| 1.25 | 2.7225 | 2.6565 | 2.7228 | 2.6563 | 2.723 | 2.656 |
| 1.30 | 3.3431 | 3.2608 | 3.3423 | 3.2614 | 3.3424 | 3.2613 |
| 1.35 | 4.0055 | 3.9084 | 4.0078 | 3.9062 | 4.0073 | 3.9067 |

Table II: Damping Length Scales

| Radial Distance in units of $R_\odot$ | Damping Length for period 60 s (in $10^8$ | Damping Length for period 60 s (in $10^8$ | Damping Length for period 180 s (in $10^8$ | Damping Length for period 180 s (in $10^8$ | Damping Length for period 300 s (in $10^8$ | Damping Length for period 300 s (in $10^8$ |
|---|---|---|---|---|---|---|

|        | cm)     | cm)     | cm)     | cm)    | cm)     | cm)     |
|--------|---------|---------|---------|--------|---------|---------|
| 1.0058 | 0.81722 | 0.83778 | 0.81747 | 0.8375 | 0.81712 | 0.83787 |
| 1.03   | 1.0388  | 1.0640  | 1.0383  | 1.0647 | 1.0382  | 1.0647  |
| 1.05   | 1.2424  | 1.2739  | 1.2425  | 1.2739 | 1.2427  | 1.2736  |
| 1.10   | 1.8364  | 1.8800  | 1.8359  | 1.8805 | 1.8348  | 1.8817  |
| 1.15   | 2.5386  | 2.6022  | 2.5392  | 2.6015 | 2.5376  | 2.6033  |
| 1.20   | 3.3402  | 3.4258  | 3.3399  | 3.4261 | 3.341   | 3.4248  |
| 1.25   | 4.2302  | 4.3352  | 4.2297  | 4.3356 | 4.2293  | 4.3361  |
| 1.30   | 5.1923  | 5.3233  | 5.1934  | 5.3222 | 5.1931  | 5.3223  |
| 1.35   | 6.2235  | 6.3781  | 6.2201  | 6.3819 | 6.2209  | 6.381   |